# Crystal structure dependent thermal conductivity in two-dimensional phononic crystal nanostructures

Junki Nakagawa[1], Yuta Kage[1], Takuma Hori[2], Junichiro Shiomi[2], and Masahiro Nomura[1,3]

[1]Institute of Industrial Science, The University of Tokyo, Tokyo 153-8505, Japan
[2]Department of Mechanical Engineering, The University of Tokyo, Tokyo, 113-8656, Japan
[3]Institute for Nano Quantum Information Electronics, The University of Tokyo, Tokyo 153-8505, Japan
*nomura@iis.u-tokyo.ac.jp

**Abstract**

Thermal phonon transport in square- and triangular-lattice Si phononic crystal (PnC) nanostructures with a period of 300 nm was investigated by measuring the thermal conductivity using micrometer-scale time-domain thermoreflectance. The placement of circular nanoholes has a strong influence on thermal conductivity when the periodicity is within the range of the thermal phonon mean free path. A staggered hole structure, i.e., a triangular lattice, has lower thermal conductivity, where the difference in thermal conductivity depends on the porosity of the structure. The largest difference in conductivity of approximately 20% was observed at a porosity of around 30%. This crystal structure dependent thermal conductivity can be understood by considering the local heat flux disorder created by a staggered hole structure. Numerical simulation using the Monte Carlo technique was also employed and also showed the lower thermal conductivity for a triangular lattice structure. Besides gaining a deeper understanding of nanoscale thermal phonon transport, this information would be useful in the design of highly efficient thermoelectric materials created by nanopatterning.

Thermal conductivity nanoengineering is one of the hottest topics in semiconductor physics. Many research studies have reported characteristic transport properties and large thermal conductivity reductions by shortening the effective mean free path (MFP) of thermal phonons in a variety of nanostructured semiconductors[1-8]. Single-crystalline Si provides an ideal phonon transport system due to its long thermal phonon MFP, which is longer than 100 nm[9,10] at room temperature, and which enables a systematic investigation of thermal conductivity control by well-defined nanopatterning, such as phononic crystal (PnC) nanostructures formed by electron beam (EB) lithography[11-18]. To date, thermal conduction in PnCs has been mainly investigated with square lattices, but there have been few experiments for other lattice types so far. Recently, Tang *et al*. performed numerical simulations and predicted that the placement of holes has a strong influence on thermal conduction[19]. Song and Chen studied thermal conductivity between PnC microstructures with a hole spacing of 4 μm aligned in a square or triangular lattices[20]. They found that strong size effects exist even in micro-sized porous Si structures at room temperature, but thermal conductivity was insensitive to the pore alignment. The pore alignment dependence of thermal conductivity is expected to be more substantial when the characteristic size of the structure approaches the scale of a thermal phonon MFP, i.e., approaches the sub-micrometer scale. A systematic experimental study of thermal conduction in Si PnC nanostructures, with dimensions well within the thermal phonon MFP range, with different lattice types and a study of the porosity dependence should provide useful information on the aforementioned physics.

In the present study, we investigated the thermal conduction in single-crystalline Si PnC nanostructures with circular holes aligned in square or triangular lattices. The period of the lattice was fixed at 300 nm, which is within the thermal phonon MFPs in Si at room temperature[10,21], in order to observe the dependence of thermal conductivity. Thermal conductivities for a variety of porosities were measured and the porosity dependence for both crystal lattices was investigated. Herein, we discuss the observed thermal conductivity measurements from the point of view of nanoscale thermal phonon transport with analyses of the experimental results and of the numerical simulation results generated via a Monte Carlo technique.

We fabricated two-dimensional (2D) PnC nanostructures with circular holes with a variety of radii ($r$s) aligned as square or triangular lattices. The PnC nanostructures were fabricated in a single-crystalline Si thin film. We used a commercially available (100) nominally boron-doped silicon-on-insulator wafer with a 145 nm-thick upper Si layer and a 1 μm-thick buried $SiO_2$ layer. The PnC nanostructures were formed via EB lithography using a reactive-ion etching inductively coupled plasma system, with $SF_6/O_2$ gas as the etchant. The buried oxide layer was removed with hydrofluoric acid in order to form the suspended structures.

Figures 1(a) and 1(b) show a scanning electron microscopy (SEM) image of the whole fabricated suspended structure with the 2D PnC structures. In the center of the structure, a 125 nm-thick Al layer was deposited to form a 4 μm × 4 μm pad on top of the Si layer, enabling the thermoreflectance measurements. The 5 μm × 5 μm central Si island was then supported by the PnC nanostructures. The 2D PnC structures were formed using circular holes aligned periodically in a square lattice and in a triangular lattice, as shown respectively in the insets in Figs. 1(a) and 1(b). The period $a$ of the PnC nanostructures was fixed at 300 nm.

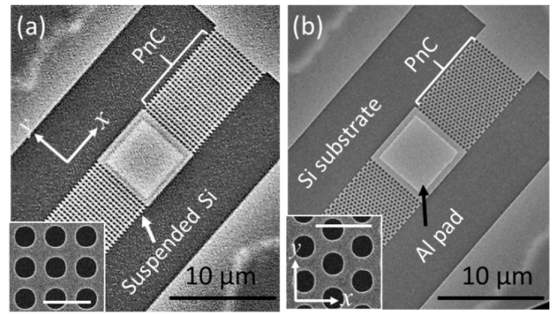

Fig. 1. SEM images of whole Si suspended structures with square-lattice 2D PnC nanostructures with a porosity of 37% (a) and triangular-lattice 2D PnC nanostructures with a porosity of 39% (b). The scale bars in the insets are 500 nm.

The thermal conductivities of the PnC nanostructures were measured using micrometer-scale time-domain thermoreflectance (μ-TDTR). The principle of this technique is the same as that of TDTR[22], but it has the benefit that it can be applied to micrometer-sized systems. The Al pad was heated by a quasicontinuous laser beam (wavelength $\lambda$ = 642 nm), with the temporal evolution of the temperature (the TDTR signal) monitored by a continuous-wave laser beam ($\lambda$ = 785 nm) as the temporal evolution of the reflectivity change ($\Delta R$). Both beams were collinearly focused on the Al pad by use of a microscope objective with a numerical aperture of 0.65, and the beam spots on the pad were approximately 700 nm in diameter. The power of the pump pulse was set so that the increase in temperature of the Al pad was less than 5 K. The radiation loss was calculated using the Stefan-Boltzmann law. The radiant energy was in the order of $10^{-4}$ of the absorbed heating pulse energy; also, was confirmed that thermal radiation was not the main heat dissipation channel under our experimental conditions[18]. All the measurements were performed in a vacuum chamber to eliminate heat-convection losses to the surroundings. The temperature evolution in the entire structure was simulated via the finite element method using COMSOL Multiphysics® software. The TDTR signals were simulated with the thermal diffusion equation. The thermal conductivities were obtained by fitting the TDTR signal by a simulated temperature evolution using the least-squares method. Detail of the measurement method and analysis of the TDTR signal can be found in ref. 18.

The thermal conductivities of the square-lattice PnC (s-PnC) and triangular-lattice PnC (t-PnC) nanostructures with various hole radii were measured. The main aim of this study is to investigate how the crystal lattice type, i.e., the difference in the placement of the holes, is reflected in the thermal conductivity. We know that phonon transport occurs in the semi-ballistic regime in the investigated PnC nanostructures from our previous investigations in similar sizes[17]. Therefore, the placement of holes was expected to have some influence on thermal conductivity.

Figures 2(a) and 2(b) show the measured TDTR signals (dots) of the s-PnCs with porosity $\Phi$ = 0 (unpatterned membrane), 19, 37, and 55%, and of the t-PnCs with $\Phi$ = 17, 39, and 56%, respectively. The measurements were performed at room temperature (295 K). The solid lines are the simulated curves using the value of thermal conductivity that gave the best fit for each PnC nanostructure at each porosity [Fig. 3(a)]. The TDTR signal increased when the pump pulse was irradiated for 500 ns and decreased as the heat was dissipated from the central Si island, which was in thermal equilibrium with the Al pad. The PnC nanostructures with higher porosities show slower decays due to possessing a lower cross-section of Si, but the thermal conductivity also decreased. This fact indicates that thermal phonons have a similar MFP to the characteristic size of the system. In this case, the characteristic size was the neck size defined by $a-2r$. The effective phonon MFP becomes shorter as the value of $\Phi$ is increased and this results in a lower thermal conductivity, which is known as the necking effect[15].

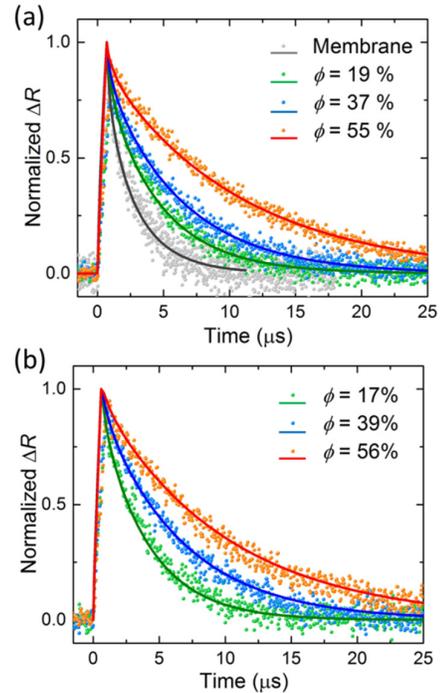

Fig. 2 Recorded TDTR signals (dots) and simulation curves (lines) for 2D PnC nanostructures with different porosities for square-lattice (a) and triangular-lattice (b) samples at room temperature.

Figure 3(a) shows the obtained thermal conductivities (dots) for the s-PnC ($\kappa_s$) and t-PnC ($\kappa_t$) nanostructures with a variety of $r$s at room temperature. The orange and blue lines are the polynomial fits to the data, which were used to calculate the difference in thermal conductivity $D(\Phi)$, as defined by:

$$D(\Phi) = \frac{\kappa_s(\Phi) - \kappa_t(\Phi)}{\kappa_s(\Phi)}. \quad (1)$$

The t-PnCs, staggered structures, have lower thermal conductivities than the s-PnCs by as much as 15-20% at the same porosity, between $\Phi$ = 20 and 50%. This result indicates that the placement of holes strongly influences heat transfer in a system with ballistic phonon transport properties. Tang *et al.* numerically studied the phonon thermal conductivity of the Si nanostructures and reported that the lower thermal conductivity of a staggered structure compared to an aligned structure. This could be explained by considering the local heat flux[19]. The staggered structure disorders the local heat flux and reduces the effective thermal phonon MFPs in the direction of the macroscopic heat flux, i.e. in the $x$ direction in Fig. 1 in our case. This explanation seems to be the main physical reason for the observed thermal conductivity difference between the two crystal lattice types. Song and Chen reported that the thermal conductivity in 7.4-μm-thick Si membranes with 4-μm-spacing and $r$ = 1 μm ($\Phi$ = 18%) is insensitive to the hole alignment[19]. However, the PnC nanostructure with 300 nm period showed nearly 20% difference. The hole placement can be reflected in thermal conductivity more strongly as the characteristic size of the PnC structure approaches the thermal phonon MFP.

Figure 3(b) shows thermal conductivities of the same samples measured at 4 K. The difference in thermal conductivities is similar to the room temperature measurement, for example, at about 17% around $\Phi$ = 35%. However, there is a clear difference in the impact of phononic nanopatterning. The measured thermal conductivities of an unpatterned thin film were 75 Wm$^{-1}$K$^{-1}$ at room temperature and 0.0488 Wm$^{-1}$K$^{-1}$ at 4 K. At $\Phi$ = 15% the thermal conductivities of the s-PnC are 56% and 28% of the values for the unpatterned thin film at room temperature and at 4 K respectively. The larger impact of phononic nanopatterning on the thermal conductivity at the lower temperature mainly stems from the difference in the thermal phonon MFP distribution. Thermal phonons have longer MFPs at low temperature, where the phonon-phonon scattering rate is much lower. Thus, the PnC nanostructures dramatically reduce the effective MFPs at 4 K, which result in a larger reduction in thermal conductivity by the patterning.

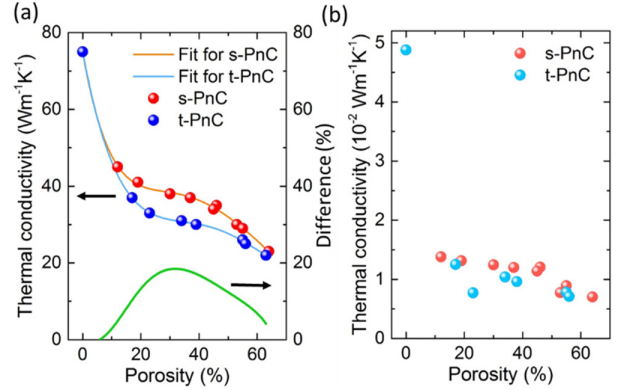

Fig. 3. Thermal conductivities of 2D PnC nanostructures with a variety of porosities measured at room temperature (a) and at 4 K (b). The orange and blue lines are the polynomial fitting curves for the measurement data (dots) for the s-PnCs and t-PnCs, respectively. The green line is the calculated curve of $D(\Phi)$.

The thermal conductivities of the PnC nanostructures were calculated to investigate whether the observed difference between the square and triangular lattice types was reproducible. In nanostructures, the thermal phonon MFPs are shortened due to phonon boundary scattering at the sidewalls of the holes. By defining the effective MFP as $\Lambda_{\text{eff}}$, the nanostructure's thermal conductivity in terms of Boltzmann transport under a relaxation time approximation (RTA) can be written as:

$$\kappa = \frac{1}{3} \sum_{\mathbf{q},s} c_{\mathbf{q},s} v_{\mathbf{q},s} \Lambda_{\text{eff},\mathbf{q},s}, \quad (2)$$

where $c$ and $v$ are the specific heat and group velocity, which depend on the phonon wave vector $\mathbf{q}$ and branch $s$. The calculation of $\Lambda_{\text{eff}}$ was realized using a Monte Carlo ray tracing simulation, and the details can be found in refs. 18 and 23. The method calculates the phonon transmission probability ($\tau$) by emitting a phonon from one side of the nanostructure with incident polar ($\theta$) and azimuthal ($\varphi$) angles, and then statistically evaluating the probability of the phonons reaching the opposite side of the simulated system. The intrinsic MPFs due to phonon-phonon scattering were set to those of bulk single-crystalline Si obtained from first principles[22]. Phonon scattering (reflection) at the surfaces was assumed to be diffusive, which should be a reasonable assumption at room temperature. $\Lambda_{\text{eff}}$ can then be obtained based on kinetic theory and Landauer's formula as:[23]

$$\Lambda_{\text{eff}} = \frac{3L}{2M} \int_0^{\pi/2} \tau(\theta,\Lambda)\cos\theta \sin\theta d\theta, \quad (3)$$

where $M$ is a modification factor of the structure accounting for the porosity of the PnC structures calculated by COMSOL Multiphysics®, and $L$ is the length of the system.

Figure 4 shows the calculated thermal conductivities in comparison with the experimental data of Fig. 3(a). The dependence of the calculated thermal conductivities on hole radius reproduces the trend seen in the experiment. Thermal conductivity decreases as the radius increases, and the reduction is steeper above $r$ = 115 nm. For both s-PnC and t-PnC nanostructures the measured thermal conductivities are lower than the simulated value. This is mainly attributed to the imperfection of the fabricated nanostructure, especially due to the surface roughness of the sidewalls of the holes. The difference in thermal conductivity between the two crystalline types is small in the calculations, but the reason for this is unclear. However, more importantly, the calculations reproduce the key structural dependence, i.e., they show that the thermal conductivity of the triangular lattice sample is lower than that of the square lattice.

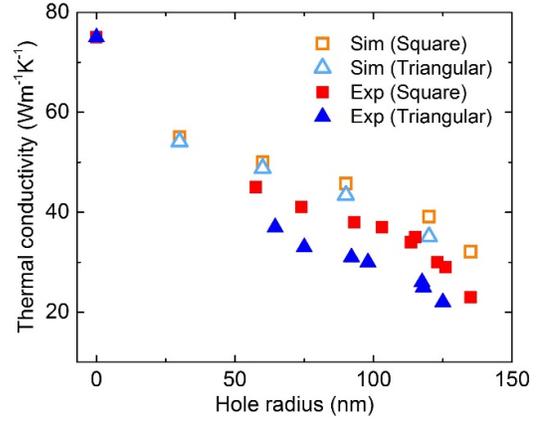

Fig. 4. Simulated and measured thermal conductivities for the s-PnC and t-PnC nanostructures with a variety of hole radii at room temperature.

In Fig. 3(a), the calculated curve $D(\Phi)$ increases as $\Phi$ increases, but declines for $\Phi$ above 35%. This is an interesting result to discuss from the viewpoint of nanoscale phonon transport. This is because the phonon scattering in the structures is dominated by surface scattering at the narrowest part, i.e., the neck of the PnC structures. The neck sizes are different for s-PnC and t-PnC nanostructures at the same porosity, as shown in Fig. 5(a). The square lattice has a smaller neck size than the triangular lattice at the same porosity. Therefore, the thermal conductivity decreases more steeply for the s-PnC samples. Figure 5(b) shows a re-plot of the thermal conductivities in Fig. 3(a), with the neck size on the horizontal axis. The difference in thermal conductivity is clearly seen at small neck sizes (high porosities) for both s-PnC and t-PnC nanostructures. We note that the thermal conductivity is greatly different, even at the same neck size. Therefore, we can conclude that the observed thermal conductivity difference can be attributed to the different placement of the holes.

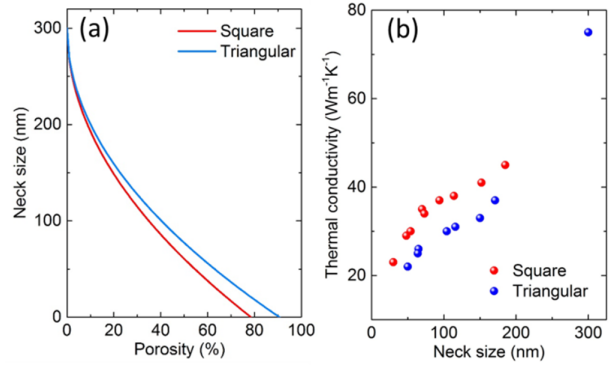

Fig. 5. (a) Calculated neck sizes for the s-PnC and t-PnC nanostructures as a function of porosity. (b) Re-plot of the thermal conductivity at room temperature plotted with neck size on the horizontal axis.

In summary, the thermal conductivities in square- and triangular-lattice PnC nanostructures with a period of 300 nm were measured, and herein have been discussed from the view point of nanoscale phonon transport. Thermal conductivity was found to differ by as much as approximately 20% at porosities between 30 and 35%. The placement of holes has a strong influence on thermal conductivity by changing the local heat flux. The difference becomes larger as the characteristic size of the structure approaches the scale of the thermal phonon MFPs. This information gives us a deeper understanding of nanoscale heat transport and would be useful in the design of highly efficient thermoelectric materials based on nanopatterning.

We thank J. Maire for the technical assistance and R. Anufriev for the fruitful discussions. This work was supported by the Project for Developing Innovation Systems of the Ministry of Education, Culture, Sports, Science and Technology (MEXT), Japan and by KAKENHI (25709090 and 15K13270).

[1] D.M. Rowe, V.S. Shukla, and N. Savvides, Nature **290**, 765 (1981).

[2] M.N. Luckyanova, J. Garg, K. Esfarjani, A. Jandl, M.T. Bulsara, A.J. Schmidt, A.J. Minnich, S. Chen, M.S. Dresselhaus, Z. Ren, E.A. Fitzgerald, and G. Chen, Science **338**, 936 (2012).

[3] C. Bera, N. Mingo, and S. Volz, Phys. Rev. Lett. **104**, 115502 (2010).

[4] T.-K. Hsiao, H.-K. Chang, S.-C. Liou, M.-W. Chu, S.-C. Lee, and C.-W. Chang, Nat. Nanotechnol. **8**, 6 (2013).

[5] J. Tang, H.-T. Wang, D.H. Lee, M. Fardy, Z. Huo, T.P. Russell, and P. Yang, Nano Lett. **10**, 4279 (2010).

[6] A.I. Hochbaum, R. Chen, R.D. Delgado, W. Liang, E.C. Garnett, M. Najarian, A. Majumdar, and P. Yang, Nature **451**, 163 (2008).

[7] J. Maire and M. Nomura, Jpn. J. Appl. Phys. **53**, 06JE09 (2014).

[8] J. Lee, J. Lim, and P. Yang, Nano Lett. **15**, 3273 (2015).

[9] A. Minnich, J. Johnson, A. Schmidt, K. Esfarjani, M. Dresselhaus, K. Nelson, and G. Chen, Phys. Rev. Lett. **107**, 1 (2011).

[10] K.T. Regner, D.P. Sellan, Z. Su, C.H. Amon, A.J.H. McGaughey, and J. a Malen, Nat. Commun. **4**, 1640 (2013).

[11] J.-K. Yu, S. Mitrovic, D. Tham, J. Varghese, and J.R. Heath, Nat. Nanotechnol. **5**, 718 (2010).

[12] P.E. Hopkins, C.M. Reinke, M.F. Su, R.H. Olsson, E. a Shaner, Z.C. Leseman, J.R. Serrano, L.M. Phinney, and I. El-Kady, Nano Lett. **11**, 107 (2011).

[13] I. El-kady, R.H.O. Iii, P.E. Hopkins, Z.C. Leseman, D.F. Goettler, B. Kim, C.M. Reinke, and M.F. Su, Prog. Rep. No. SAND2012-0127 (Sandia Natl. Lab. CA) (2012).

[14] E. Dechaumphai and R. Chen, J. Appl. Phys. **111**, 073508 (2012).

[15] A. Jain, Y.-J. Yu, and A.J.H. McGaughey, Phys. Rev. B **87**, 195301 (2013).

[16] N. Zen, T. a. Puurtinen, T.J. Isotalo, S. Chaudhuri, and I.J. Maasilta, Nat. Commun. **5**, 1 (2014).

[17] M. Nomura, J. Nakagawa, Y. Kage, J. Maire, D. Moser, and O. Paul, Appl. Phys. Lett. **106**, 143102 (2015).

[18] M. Nomura, Y. Kage, J. Nakagawa, T. Hori, J. Maire, J. Shiomi, R. Anufriev, D. Moser, and O. Paul, Phys. Rev. B **91**, 205422 (2015).

[19] G.H. Tang, C. Bi, and B. Fu, J. Appl. Phys. **114**, 184302 (2013).

[20] D. Song and G. Chen, Appl. Phys. Lett. **84**, 687 (2004).

[21] K. Esfarjani and G. Chen, Phys. Rev. B **84**, 085204 (2011).

[22] D.G. Cahill, Rev. Sci. Instrum. **75**, 5119 (2004).

[23] T. Hori, J. Shiomi, and C. Dames, Appl. Phys. Lett. **106**, 171901 (2015).